\documentclass[a4paper]{jpconf}

\usepackage{float,amsmath,amssymb,amsfonts,mathtools}
\usepackage{graphicx,fancyhdr,multirow,hyperref}
\usepackage[a4paper]{geometry}
\usepackage{textcomp}
\usepackage{caption}
\usepackage{subcaption}

\usepackage[comma,authoryear]{natbib}
\newcommand{\Tab}{Table~}
\newcommand{\Sec}{Section~}

\makeatletter
\newcommand{\eq}[1]{
    \@ifnextchar\bgroup{\ifhmode \ifnum\spacefactor>2999 Equations\else Eqs.\fi\else Equations\fi ~{(#1)}\checknextarg}{\ifhmode \ifnum\spacefactor>2999 Equation\else Eq.\fi\else Equation\fi ~{(#1)}\spacevalue}}
\newcommand{\checknextarg}{\@ifnextchar\bgroup{\gobblenextarg}{ }}
\newcommand{\gobblenextarg}[1]{\@ifnextchar\bgroup{, (#1) \gobblenextarg}{ and (#1)\spacevalue}}
\newcommand{\spacevalue}{\@ifnextchar.{}{\@ifnextchar,{}{ }}}
\makeatother

\makeatletter
\newcommand{\figs}[1]{
    \@ifnextchar\bgroup{\ifhmode \ifnum\spacefactor>2999 Figures\else Figs.\fi\else Figures\fi ~{#1}\checknextargfig}{\ifhmode \ifnum\spacefactor>2999 Figure\else Fig.\fi\else Figure\fi ~{#1}\spacevaluefig}}
\newcommand{\checknextargfig}{\@ifnextchar\bgroup{\gobblenextargfig}{ }}
\newcommand{\gobblenextargfig}[1]{\@ifnextchar\bgroup{, #1 \gobblenextargfig}{\consecutivefigend{#1}\spacevaluefig}}
\newcommand{\consecutivefigend}[1]{\if\relax\detokenize{#1}\relax{}\else{ and #1}\fi}
\newcommand{\spacevaluefig}{\@ifnextchar.{}{\@ifnextchar,{}{ }}}
\makeatother

\def\mpa{{\mathrm{MPa}}}
\def\mhz{{\mathrm{MHz}}}

\def\cald{{Calder{\'o}n}~}
\def\pd{{Pad{\'e}}~}

\def\ie{{i.e.~}}

\def\ltwonorm{{$l_{\scriptscriptstyle 2}-\mathrm{norm}~$}}

\begin{document}

\title[]{A fast full-wave solver for trans-abdominal high-intensity focused ultrasound treatment planning}

\author{S. R. Haqshenas$^{1,2}$, P. G\'elat$^1$, E. van {'}t Wout$^3$, T. Betcke$^2$, N. Saffari$^1$}
\address{$ˆ1$ Department of Mechanical Engineering, University College London, London, WC1E 7JE, UK}
\address{$ˆ2$ Department of Mathematics, University College London, London, WC1H 0AY, UK} 
\address{$ˆ3$ Institute for Mathematical and Computational Engineering, Pontificia Universidad Cat{\'o}lica de Chile, Santiago, Chile}

\ead{s.haqshenas@ucl.ac.uk}

\begin{abstract}

High-intensity focused ultrasound (HIFU) is a promising treatment modality for the non-invasive ablation of pathogenic tissue in many organs. Optimal treatment planning strategies based on high-performance computing methods are expected to form a vital component of a successful clinical outcome in which healthy tissue is preserved and accurate focusing achieved. The practical application of the algorithms informing these strategies depends on the ability to produce fast and accurate full-wave patient-specific simulations, with minimal computational overheads. For realistic clinical scenarios, all simulation methods which employ volumetric meshes require several hours or days to run on a computer cluster. The boundary element method (BEM) is an effective approach for modelling the wave field because only the boundaries of the hard and soft tissue regions require discretisation. A multiple-domain BEM formulation with a novel preconditioner for solving the Helmholtz transmission problem (HTP) is presented in this paper. Numerical experiments are performed to solve HTP in multiple domains comprising: i) human ribs, an abdominal fat layer and liver tissue, ii) a human kidney and a perinephric fat layer, exposed to the acoustic field generated by a HIFU array transducer. The time required to solve the equations associated with these problems on a single workstation is of the order of minutes. These results demonstrate the great potential of this new BEM formulation for rapid and accurate HIFU treatment planning.

\end{abstract}


\section{Introduction}
Surgical and ablative techniques are the most effective local therapies for solid malignancies \citep{cranston2015review}. The significant side effects associated with surgical interventions have led to an ongoing quest for safer, more efficient and better tolerated alternatives. In recent years, there has been a notable shift away from open surgery towards less invasive procedures such as laparoscopic and robotic surgery, and from there to other methods for \textit{in situ} tumour destruction, often involving energy based destruction. These include embolization \citep{waked2017transarterial}, radiofrequency \citep{crocetti2018radiofrequency}, microwave \citep{meloni2017microwave} and laser ablation \citep{sartori2017laser}, cryoablation \citep{werner2018single} and high-intensity focused ultrasound (HIFU) \citep{ter2007therapeutic}. HIFU is a procedure which uses high-intensity ultrasound to thermally ablate a localised region of tissue. For abdominal applications, the ultrasound is typically generated by a focused transducer located extracorporeally.

HIFU treatment planning may be defined as obtaining a solution to a mathematical inverse problem whose output provides a patient specific and clinically suitable means of thermally ablating a volume of diseased tissue, whilst keeping the energy absorption in healthy tissue below an ablative threshold. There is growing clinical consensus that, for an optimal outcome, the planning phase must be achieved prior to the intervention, using numerical simulations on anatomical data obtained from patient scans. Optimal excitation protocols of a multi-element transducer may then be determined \citep{gelat2012}, thus ensuring that diseased tissue is accurately ablated. Indeed, the treatment of abdominal tumours presents challenges due to soft tissue heterogeneity and the presence of strong scatterers, such as bone and the bowel. In HIFU ablation of tumours of the liver, ribs both scatter and absorb ultrasound, which may lead not only to an aberration of the focus but can also cause skin burns \citep{casper2010optimal,gelat2014,ramaekers2017improved}.

The most promising full-wave computationally efficient 3D numerical method for HIFU treatment planning of transcostal tumours is currently the boundary element method (BEM) \citep{elwin2015,betcke2017}. In BEM, the partial differential equation that models the wave propagation is essentially reformulated into a boundary integral equation defined on the interfaces. The boundary solutions and potential integrals are used to calculate the field at any point in the domain. Being based on Green$^{\scriptscriptstyle {'}}$s function representations, BEM is almost devoid of the numerical dispersion and dissipation effects commonly associated with numerical schemes such as $k$-space pseudospectral and finite-difference time domain methods \citep{marburg2018,treeby2014}. For sound-hard scatterers, issues of long computing times have been overcome through analytical preconditioners used in conjunction with hierarchical matrix ($\mathcal{H}$-matrix) compression techniques \citep{betcke2017}. 

In order for such schemes to be incorporated into a clinically-relevant treatment planning framework, it is vital that they be extended to include the ability to deal with some degree of tissue heterogeneity. Whilst soft tissue such as skin, muscle and fat have similar compressional wave speeds and densities \citep{duck2013physical}, fat attenuates ultrasound energy significantly and causes beam aberration at the focus. This impedes focal heating during HIFU treatment of renal cancer, due to the presence of perinephric fat \citep{ritchie2013attenuation,cranston2015review,suomi2018full}. It also remains to be established to which extent ultrasound can penetrate rib bone and how important it is to account for this phenomenon when modelling the propagation of ultrasound in the abdominal region. The ability to model the propagation of ultrasound through ribs and assess the importance of this relative to the case of locally reacting ribs \citep{gelat2014} would be beneficial, particularly when studying the heat transfer mechanisms which lead to skin burns in HIFU patients \citep{ramaekers2017improved}. There is therefore a requirement for resorting to multi-domain BEM to deal with the treatment planning complexities outlined above.

A number of multi-domain acoustic BEM formulations have been proposed \citep{cheng1991multidomain,medeiros2014predicting,huang2016new}. In these approaches, the boundary integral equation (BIE) was applied to each domain. Due to the matrices resulting from domain discretisation being fully populated, the set-up time and memory consumption scale quadratically with respect to the number of degrees of freedom \citep{brunner2010comparison}. Fast BEM schemes have been introduced for the Helmholtz equation, resulting in a quasilinear complexity. The merits of such methods, namely the fast multipole method and hierarchical matrices, were discussed by Brunner \textit{et al.} \citep{brunner2010comparison} where a scattering problem involving a 22 m long, 2 m thick cylinder immersed in water was analyzed for excitation frequencies up to 200 Hz. Such fast BEM schemes have been applied to multi-domain Helmholtz problems. A fast multipole BEM method for 3D multi-domain acoustic scattering problems was described by Wu et al \citep{wu2012fast}.

The above studies investigate cases where the \textit{ka} (where \emph{k} is the wavenumber and \emph{a} is the characteristic size of the domain) is substantially less than that encountered in HIFU applications. Determining which BEM scheme is optimal for solving multi-domain Helmholtz problems at frequencies relevant to HIFU problems remains an active area of research. Previously, a fast BEM model was developed to solve the scattering from a single and perfectly rigid domain  \citep{elwin2015,betcke2017}. In this paper, we present a fast multiple domain BEM model to compute both the scattered and transmitted ultrasonic fields in soft-tissue and bone. This formulation employs advanced preconditioning and $\mathcal{H}$-matrix compression techniques which enable the wave equation to be solved in large domain sizes relative to the wavelengths involved. The equations were implemented and solved using the open source library Bempp \citep{smigaj2015}. The simulations were validated against analytical solutions, where available. Subsequently, the scattered field was calculated for the following computational domains: i) a layer of abdominal fat, human ribs and liver tissue, and ii) a human kidney model and a perinephric fat layer. In both scenarios, the incident field is generated by a spherical section array transducer operating at the frequency of $1~\mhz$. Finally, the suitability and limitations of this BEM scheme for use in HIFU treatment planning are discussed.

\section{Model formulation}
In this paper, we consider the problem of solving acoustic wave propagation in a medium made of the exterior domain $\Omega^0$ and non-overlapping domains $\Omega^j$, $~j=1,\dots,n$ with different physical properties, i.e. density and wave speed. A single trace formulation which involves one Dirichlet data and one Neumann data at each point of each interface is used and implemented. As a result, this formulation holds true where only two domains are in contact with each other, thus not allowing for triple points.

\subsection{Wave equation}
It is well known that the propagation of ultrasonic waves in tissue at amplitudes relevant to HIFU is nonlinear, where nonlinearities are mainly confined to the focal region of the field \citep{yuldashev2013}. Crucial information pertaining to treatment planning of cancers of the abdomen can however be obtained via linear models. Suomi et al \citep{suomi2018full} report substantial focal splitting when planning treatment for cancers of the kidney, where acoustic full-wave simulations on a domain size of $200 \times 200 \times 200~\mathrm{mm}^3$ using a $k$-space pseudospectral method required approximately two days to complete on a 400 core cluster. Compensating for beam aberrations is likely to involve techniques employing a full matrix capture of the transducer array, thus necessitating as many forward simulations as there are elements of the array (typically 256, 512 or 1024). The forward calculations may then be incorporated into a constrained optimisation scheme \citep{gelat2012,gelat2014} so that field quantities are kept below a specified threshold. Fast linear full-wave modelling schemes applied to heterogeneous media therefore have a distinct advantage in compensating for soft tissue and bone heterogeneities. The first phase of a treatment plan can be envisaged as consisting of a purely linear computation to calculate the optimal array excitation vector for sparing healthy tissue. This excitation vector can then serve as input data into a nonlinear model confined to the focal region so that precise thermal dose metrics may be more accurately determined.

\begin{figure}[bh]
    \begin{center}
        \includegraphics{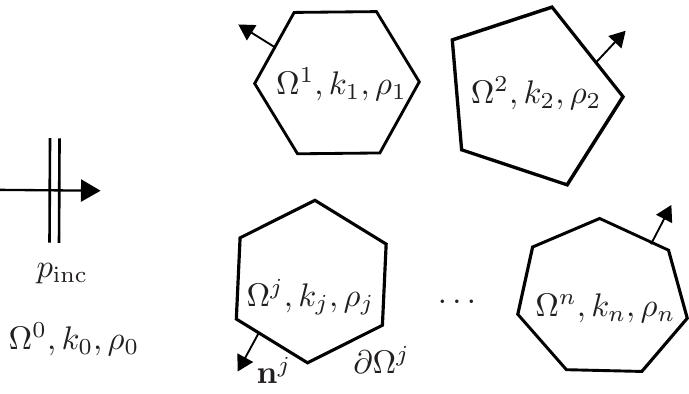}
    \end{center}
    \caption{Schematic diagram of the multiple domains. Refer to the text for the details.}
    \label{fig:domains_diagram}
\end{figure}

The linear full-wave propagation is modelled by the Helmholtz system as follows

\begin{align}
    \label{eq:model}
    \left\{
    \begin{tabular}{p{0.7\columnwidth}p{0.3\columnwidth}}
        $\Delta p^j(\mathbf{x}) + k_j^2 ~p^j(\mathbf{x}) ~= 0$, & $\text{for}~\mathbf{x}\in \Omega^j$ \\
        $\Delta p_s(\mathbf{x}) + k_0^2 ~p_s(\mathbf{x}) ~= 0$, &$\text{for}~\mathbf{x}\in \Omega^0$ \\
        $\gamma_D^{j,+} \left(p_s(\mathbf{x})~+~p_{\mathrm{inc}}(\mathbf{x}) \right)~=~ \gamma_D^{j,-} p^j(\mathbf{x}) $, & \\
          & $\text{for}~\mathbf{x}\in \partial\Omega^j$ \\ 
        $\dfrac{1}{\rho_0} \gamma_N^{j,+} \left(p_s(\mathbf{x}) + p_{\mathrm{inc}}(\mathbf{x}) \right) ~=~ \dfrac{1}{\rho_j}\gamma_N^{j,-} p^j(\mathbf{x}) $, & \\
         & $\text{for}~\mathbf{x}\in \partial\Omega^j $\\
        $\lim_{ \left|\mathbf{x}\right| \to \infty} \left|\mathbf{x}\right| \left( \nabla p_s(\mathbf{x}) \cdot \dfrac{\mathbf{x}}{\left|\mathbf{x}\right|}~-~ik_0 p_s(\mathbf{x}) \right)~=~0. $& 
    \end{tabular}
    \right.
\end{align}

\noindent where $k_j$, $\rho_j$, $\gamma_D^{j,\pm}$ and $\gamma_N^{j,\pm}$ are the wavenumber, the mass density, and the Dirichlet and Neumann traces associated with the domain $j$ for $j=1,...,n$ taken from the exterior(+)/interior(--), respectively. The wavenumber $k_j$ is complex with the positive real part equals to $2 \pi/\lambda$ where $\lambda$ is the wavelength in meters, and the positive imaginary part being $\alpha_0 f^b$ where $\alpha_0$ is the attenuation coefficient in $\mathrm{Neper~m^{-1}~Hz^{-1}}$, $f$ is the frequency in Hz and $b$ is the exponent. This frequency power law model is used to account for the attenuation of wave propagation in soft and hard tissues \citep{hamilton1998}. The traces are defined as follows:

\begin{align}
    \label{eq:traces}
    \begin{cases}
        \gamma_D^{j,\pm} p(\mathbf{x}) ~:=~ \lim_{\Omega^{\pm} \ni \mathbf{x^\prime} \to \mathbf{x}} p(\mathbf{x^\prime}) \\
        \gamma_N^{j,\pm} p(\mathbf{x}) ~:=~ \lim_{\Omega^{\pm} \ni \mathbf{x^\prime} \to \mathbf{x}} \nabla p(\mathbf{x^\prime}) \cdot \mathbf{n}^j(\mathbf{x^\prime}) \\
    \end{cases}
\end{align}

\noindent where $\mathbf{n}^j(\mathbf{x})$ is the normal vector on the boundary $\partial \Omega^j$ and points outwards to the exterior of domain $\Omega^j$ (for both $\gamma_N^{j,\pm}$ traces), see \figs{\ref{fig:domains_diagram}}. Here, $p_s(\mathbf{x})$, $p_{\mathrm{inc}}(\mathbf{x})$, are the \emph{scattered} and \emph{incident} pressure fields in the exterior, $p^j(\mathbf{x})$ is the pressure field in $\Omega^j$. For the sake of simplicity, the spatial variable $\mathbf{x}$ will be dropped from now on.

The last line in \eq{\ref{eq:model}} is the Sommerfeld radiation condition that requires the scattered waves to be outgoing in the unbounded domain $\Omega^0$.

\subsection{Coupled boundary integral formulation}

The \cald representation of the Helmholtz problem in the exterior domain $\Omega^0$ is given by, see \ref{sec:CaldOp} for the derivations,

\begin{align}
    \label{eq:Form1_calderon_sys_ext_ptot}
   \left( \frac{1}{2}\mathsf{Id}~-~\mathsf{A}_{k_0} \right)~\gamma^{+}p_{\mathrm{tot}}~=~\gamma^{+}p_s,
\end{align}

\noindent where $\gamma^{+}:= \left[\gamma_D^{+} ~~ \gamma_N^{+} \right]^T$ is the exterior \emph{Cauchy trace operator} at $\bigcup_{j=1}^n \partial\Omega^j$, $\mathsf{Id}$ is the identity operator, $\mathsf{A}_{k_0}$ is the \cald operator evaluated with the wavenumber of the exterior domain, $p_{\mathrm{tot}}~=~p_{\mathrm{inc}}~+~p_s$ is the total pressure in the exterior domain $\Omega^0$. Restricting the boundary integral operators to a single domain $j$ results in

\begin{align}
    \label{eq:Form1_calderon_sys_ext_ptot_domain_j}
   & \left( \frac{1}{2}\mathsf{Id}~-~\mathsf{A}^j_{k_0} \right)~\gamma^{j,+}p_{\mathrm{tot}}~-~\sum_{i \neq j}\mathsf{A}_{k_0}^{ji}\gamma^{i,+} p_{\mathrm{tot}}  \nonumber \\
   &~=~\gamma^{j,+}p_s.
\end{align}

Here, $\mathsf{A}_{k_0}^{ji}$ is the \cald operator that takes the wavenumber $k_0$ and maps a surface potential at $\partial\Omega^i$ to $\partial\Omega^j$. The term $\sum_{i \neq j}\mathsf{A}_{k_0}^{ji}\gamma^{i,+} p_{\mathrm{tot}}$ accounts for all cross scatterings between domains $\Omega^i$ and $\Omega^j$ for $i=1,\dots,n,~ i \neq j$.  

For the interior of domain $\Omega^j$, the \cald representation of the Helmholtz problem reads, see \ref{sec:CaldOp} for the derivations,

\begin{align}
    \label{eq:Form1_calderon_sys_int}
   \left( \frac{1}{2}\mathsf{Id}~+~\mathsf{A}^j_{k_j} \right)~\gamma^{j,-}p^j~=~\gamma^{j,-}p^j,
\end{align}

\noindent where $\mathsf{A}^j_{k_j}$ is the \cald operator evaluated in the interior of domain $\Omega^j$.

The boundary conditions shown in \eq{\ref{eq:model}} can be represented as follows

\begin{subequations}
\label{eq:Form1_BCs}
\begin{align}
& \gamma_D^{j,-}p^j~=~\gamma_D^{j,+}p_{\mathrm{tot}}, \\
& \gamma_N^{j,-}p^j~=~\dfrac{\rho_j}{\rho_0}\gamma_N^{j,+}p_{\mathrm{tot}}.
\end{align}
\end{subequations}

\noindent Defining $D_j~:=~\begin{bmatrix}\mathsf{Id} & \mathsf{0}\\\mathsf{0} & \rho_j / \rho_0 \end{bmatrix}$, \eq{\ref{eq:Form1_BCs}} transforms to $\gamma^{j,-}p^j~=~D_j \gamma^{j,+}p_{\mathrm{tot}}$. Substituting this relationship into \eq{\ref{eq:Form1_calderon_sys_int}} and multiplying both sides with ${D_j}^{-1}$ produces

\begin{align}
    \label{eq:Form1_calderon_sys_int_ptot}
    \left( \frac{1}{2}\mathsf{Id}~+~\widetilde{\mathsf{A}}^j_{k_j}\right)~\gamma^{j,+}p_{\mathrm{tot}}~=~\gamma^{j,+}p_{\mathrm{tot}},
\end{align}
    
\noindent where $\widetilde{\mathsf{A}}_{k_j}^j~:=~{D_j}^{-1}\mathsf{A}_{k_j}^j~D_j$.

\noindent Subtracting \eq{\ref{eq:Form1_calderon_sys_ext_ptot_domain_j}} from \eq{\ref{eq:Form1_calderon_sys_int_ptot}} produces

\begin{align}
    \label{eq:Form2_calderon_sys_simp2}
        & \left( \mathsf{A}^j_{k_0}~+~\widetilde{\mathsf{A}}^j_{k_j} \right)\gamma^{j,+}p_{\mathrm{tot}}~+~\sum_{i \neq j} \mathsf{A}_{k_0}^{ji}\gamma^{i,+}p_{\mathrm{tot}} \nonumber \\
        &~~=~\gamma^{j,+} p_{\mathrm{tot}}~-~\gamma^{j,+} p_s , 
    \end{align}

\noindent which rearranges to the final form of the formulation as follows

\begin{align}
\label{eq:Form2_calderon_sys_simp3}
    & \left( \mathsf{A}^j_{k_0}~+~\widetilde{\mathsf{A}}^j_{k_j} \right)\gamma^{j,+}p_{\mathrm{tot}}~+~\sum_{i \neq j} \mathsf{A}_{k_0}^{ji}\gamma^{i,+}p_{\mathrm{tot}} \nonumber \\
    &~~=~\gamma^{j,+} p_{\mathrm{inc}} \hspace{0.3\columnwidth} \forall j=1,\dots,n.
\end{align}

This equation is called the Poggio-Miller-Chan-Harrington-Wu-Tsai (PMCHWT) formulation for the multiple domain HTP. This represents a set of linear equations which can be condensed into a block system $\widehat{\mathsf{A}}_k\mathsf{u}~=~\mathsf{b}$ where

\begin{align}
    \label{eq:Form2_matrix_format_p1}
    \widehat{\mathsf{A}}_k:=~\left[\widehat{\mathsf{A}}_k\right]_{ji}~=~
    \begin{cases}
        \mathsf{A}_{k_0}^j~+~\widetilde{\mathsf{A}}_{k_j}^j & j=i\\
        \mathsf{A}_{k_0}^{ji}  & j \neq i \\
    \end{cases},
\end{align}

\noindent and

\begin{align}
    \label{eq:Form2_matrix_format_p2}
    \mathsf{u}_j~=~\gamma^{j,+}p_{\mathrm{tot}}, \quad \mathsf{b}_j~=~\gamma^{j,+} p_{\mathrm{inc}}.
\end{align}

\noindent In the above equations the \cald operator is a $2\times2$ block matrix represented by

\begin{align}
\mathsf{A}_{k_j}^j~:=~\begin{bmatrix}
-~\mathsf{K}^j  & \mathsf{V}^j  \\
\mathsf{W}^j &  \mathsf{K^{'}}^j
\end{bmatrix},
\end{align}

\noindent where $\mathsf{V}^j$, $\mathsf{K}^j$, $\mathsf{K^{'}}^j$, $\mathsf{W}^j$ are the single layer, double layer, adjoint double layer, and hypersingular boundary integral operators, respectively, see \ref{sec:CaldOp} for their definitions. Thus, $4n(n+1)$ boundary integral operators should be evaluated to construct the PMCHWT formulation for the HTP for $n$ domains.

\subsection{Operator preconditioning}

The Galerkin method is used for the discretisation of boundary integral operators. The boundary $\partial \Omega^j$ is meshed by triangular elements. The discrete formulation results in a linear system of equations which corresponds to the matrix problem $\mathbf{A}\mathbf{u}~=~\mathbf{b}$ where

\begin{align}
\label{eq:Galerkin_disc_lhs}
\left[\mathbf{A}\right]_{rq}~=~\left< \widehat{\mathsf{A}}_k\eta_q,\eta_r \right>, \quad \forall q,r= 1,2,\dots,z
\end{align}

\noindent $\mathbf{u}=[u_1, u_2, \dots, u_z]$ is the vector of unknown solution of the variational problem and $\mathbf{b}=[b_1, b_2, \dots, b_z]$ with

\begin{align}
\label{eq:Galerkin_disc_rhs1}
b_{r}~=~\left< \gamma^{j,+} p_{\mathrm{inc}},\eta_r \right>.
\end{align}

Here, $z$ is the number of degrees of freedom (dofs), $\eta_q$ and $\eta_r$ are the trial and test functions, and $\left<.,.\right>$ is the standard $L^2$ dual pairing. The test and basis functions are linear piecewise continuous functions on the triangular surface mesh. The reader is referred to \citep{smigaj2015,betcke2017} for further details.

This linear discrete system of equations is usually ill-conditioned at high driving frequencies and acoustic impedance mismatch between the exterior and interior domains. Consequently, iterative solvers such as GMRES converge very slowly. Hence, to improve the convergence rate of the iterative solvers. the preconditioned continuous problem $\mathsf{C}^{-1}\widehat{\mathsf{A}}_k\mathsf{u}~=~\mathsf{C}^{-1}\mathsf{b}$ is solved instead where the operator $\mathsf{C}$ is defined such that the spectral condition number of the preconditioned system is smaller than the original system. The operator preconditioners are designed using the mapping properties of the BIE operators -- instead of the discretised matrix -- which allows them to be easily combined with acceleration and compression algorithms \citep{hiptmair2006, kirby2010}.

The performance of conventional operator preconditioners of PMCHWT formulation substantially deteriorates in scenarios with large wavenumbers \citep{antoine2016} and high contrast domains \citep{gossye2018}. Such scenarios are common in HIFU treatment planning applications. Therefore, a new block-diagonal preconditioner was developed to regularise the PMCHWT equations for the multiple domain Helmholtz transmission problem. This new preconditioner is based on the Neumann-to-Dirichlet, denoted by $\Lambda_\mathrm{NtD}$, and Dirichlet-to-Neumann maps, $\Lambda_\mathrm{DtN}$. $\Lambda_\mathrm{NtD}$ maps the normal gradient of the pressure field on the boundary $\partial \Omega^j$ to the pressure field on the surface. $\Lambda_\mathrm{DtN}$ performs the inverse map. These maps evaluated in the exterior domain satisfy the following relationships \citep{antoine2007,darbas2013}

\begin{align}
\label{eq:OSRC_maps_properties}
\Lambda^{j,+}_\mathrm{DtN} ~\mathsf{V}^{j,+}~=~-\frac12 \mathsf{I}~+~\mathsf{K^\prime}^{j,+}, \\ \nonumber
\Lambda^{j,+}_\mathrm{NtD} ~\mathsf{W}^{j,+}~=~-\frac12 \mathsf{I}~-~\mathsf{K}^{j,+}.
\end{align}

The double layer and adjoint double layer operators are compact and therefore $\Lambda^{j,+}_\mathrm{DtN} ~\mathsf{V}^{j,+}$ and $\Lambda^{j,+}_\mathrm{NtD} ~\mathsf{W}^{j,+}$ are better conditioned than single layer and hypersingular operators. Thus, the DtN and NtD maps are used to build the following block diagonal preconditioner

\begin{align}
\label{eq:OSRC_prec}
\mathsf{C}^{-1}~:=~\left[\mathsf{C}\right]_{ji}~=~
\begin{cases}
\mathsf{C}^j & j=i\\
0  & j \neq i \\
\end{cases},
\end{align}

where

\begin{align}
\label{eq:OSRC_prec_blocks}
\mathsf{C}^j~:=~\begin{bmatrix}
\Lambda_\mathrm{DtN}^{j,+} & \\
  &  \Lambda_\mathrm{NtD}^{j,+}
\end{bmatrix}.
\end{align}

\noindent The preconditioner is designed with information of the self-interaction of the domains only. This is because the conditioning of $\widehat{\mathsf{A}}_k$ is mainly influenced by the diagonal blocks rather than cross-interactions between domains (the off-diagonal blocks).

In order to use this preconditioner with the PMCHWT formulation, the \cald operators should be vertically permuted to place the hypersingular and single layer operators on the main diagonal. To preserve the RHS of \eq{\ref{eq:Form2_calderon_sys_simp3}}, the blocks of the unknown vector $\mathbf{u}_j = \left[\gamma_D^{j,+} p_{\mathrm{tot}} ~~ \gamma_N^{j,+} p_{\mathrm{tot}} \right]^T$ are permuted as well.

To implement this preconditioner, we need to calculate the DtN and NtD maps. These are non-local pseudo-differential operators with no closed-form solutions. However, an accurate approximation based on the high frequency asymptotics of them can be achieved using the on-surface radiation condition (OSRC) method \citep{antoine1999,antoine20082}. The expressions of the NtD and DtN maps approximated by the OSRC approach are given by

\begin{subequations}
\label{eq:OSRC_DtN_NtD_maps}
\begin{align}
& \widetilde{\Lambda_\mathrm{DtN}}~=~ik~\sqrt{~1~+~\dfrac{\Delta_\Gamma}{k_\epsilon^2} }, \label{eq:OSRC_Dtn} \\
& \widetilde{\Lambda_\mathrm{NtD}}~=~\dfrac{1}{ik}~\left(~1~+~\dfrac{\Delta_\Gamma}{k_\epsilon^2}\right)^{-1/2}. \label{eq:OSRC_NtD}
\end{align}
\end{subequations}

\noindent where $k_\epsilon~=~k~+~i\epsilon$ where $\epsilon>0$ is a damped wavenumber and $\Delta_\Gamma$ denotes the Laplace-Beltrami operator. The singularities are avoided by using the damped wavenumber. A good choice of damping is $\epsilon~=~0.4k^{-1/3}R_\epsilon^{-2/3}$ with $R_\epsilon$ the radius of the scatterer. The square root operations in above equations are approximated with a \pd series expansion with $N_{\mathrm{Pade}}$ terms. The reader is referred to \citep{elwin2015,betcke2017,darbas2013} for the details. Since the OSRC approximations of the NtD and DtN maps, \eq{\ref{eq:OSRC_Dtn}}{\ref{eq:OSRC_NtD}}, are used to compute the elements of the preconditioner \eq{\ref{eq:OSRC_prec}}, it will be referred to as the OSRC block diagonal preconditioner. It should be noted that this block diagonal OSRC preconditioner is sparse which results in fast computation.

\section{Numerical experiments}

The number of surface elements required by BEM to represent the wave propagation is of $\mathcal{O}(k^2)$ which results in the memory footprint of $\mathcal{O}(k^4)$, as Galerkin discretisation of the boundary integral operators results in dense matrices. This prohibits the implementation of this BEM formulation for HIFU treatment planning on a typical computational platform. To alleviate this problem, the hierarchical matrix compression with the tolerance set to $10^{-6}$ is used to assemble the discretised boundary integral operators in a compressed format. Refer to \citep{elwin2015,betcke2017} for complementary information about hierarchical matrices and their application for solving Helmholtz problems.

A workstation with 32 processors (Intel\textsuperscript{\tiny{\textregistered}} Xeon(R) CPU E5-2683 v4{@}2.10 GHz) and 512 GB RAM was used for performing the following numerical experiments. First, the preconditioned formulation and its implementation was validated against analytical solutions for a model problem consisting of a spherical object in \Sec \ref{sec:model_problem}. Subsequently, the scattered field from i) human ribs and an abdominal fat layer, ii) a human kidney and a perinephric fat layer, exposed to the acoustic field generated by a HIFU array transducer is computed and discussed in \Sec\ref{sec:hifu_ribs_kidney}.

The presented formulation was implemented in Python using the open source Bempp library \citep{smigaj2015}, version 3.3.4. The GMRES algorithm of the SciPy library with the restart parameter set to $100$ was used for solving the discrete system of equations. The GMRES scheme terminates when the relative residual is below a specified relative tolerance, which was set to $10^{-5}$. The efficiency of the OSRC preconditioner \eq{\ref{eq:OSRC_DtN_NtD_maps}} is affected by parameters $R_\epsilon$ and $N_{\mathrm{Pade}}$. Although one can choose different $R_\epsilon$ and $N_{\mathrm{Pade}}$ for different domains, one set of values were selected to construct the preconditioner for the sake of simplicity. The larger $N_{\mathrm{Pade}}$ the more accurate the approximation of DtN and NtD maps are, but at the cost of more numerical operations at each iteration. The numerical investigations demonstrate the following points: i) a relatively small size of \pd series, 4 for spheres and 8 for anatomical meshes, is sufficient to calculate the OSRC preconditioner and achieve expected performance, ii) setting $R_\epsilon$ to a value of one order of magnitude smaller than the characteristic size of the largest domain in the model leads to a fast convergence.

\begin{figure}[t!]
    \begin{center}
    \includegraphics{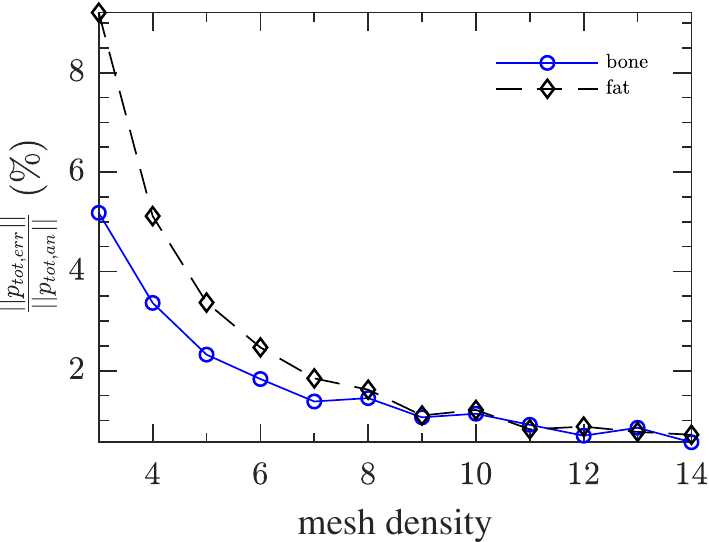}
    \end{center}
    \caption{The percentage of the \ltwonorm of the difference between the computed total pressure by BEM and the analytical total pressure at different mesh densities. }
    \label{fig:norms_error}
\end{figure}

\subsection{Single sphere model}\label{sec:model_problem}
The HTP problem of a single sphere object impinged by a plane wave is considered. The analytical solution is obtained using the classical series expansion method for a penetrable acoustic sphere \citep{anderson1950,morse1968}. The driving frequency is considered to be $1~\mhz$ which is typically used in HIFU applications for the liver and kidney. The exterior medium is assumed to be water and the sphere possesses the physical properties of i) rib bones and ii) abdominal fat. The physical properties are listed in \Tab \ref{tab:physical_properties}. The diameter of the sphere is 1 cm which is large enough to include multiple interior reflections at the wavelength of the sphere material, and which is of the order of magnitude of the anatomical heterogeneities considered in \Sec\ref{sec:hifu_ribs_kidney} of this paper. The problem was solved for a range of mesh densities, the latter parameter corresponding to the number of elements per wavelength associated with the material which has the smallest speed of sound. The value of $R_\epsilon=5\times10^{-4}~\mathrm{m}$ is used for all the sphere simulations.

\begin{figure}
    \begin{subfigure}[b]{0.45\columnwidth}
        \begin{center}
            \includegraphics{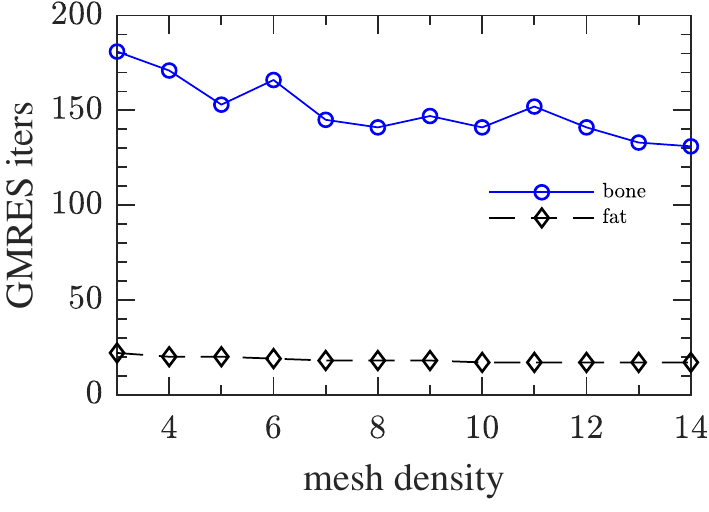}
        \end{center}
        \subcaption{Number of GMRES iterations at different mesh densities.}
        \label{fig:gmres_iters}
    \end{subfigure}
    \hspace{5mm}
    \begin{subfigure}[b]{0.45\columnwidth}
        \begin{center}
            \includegraphics{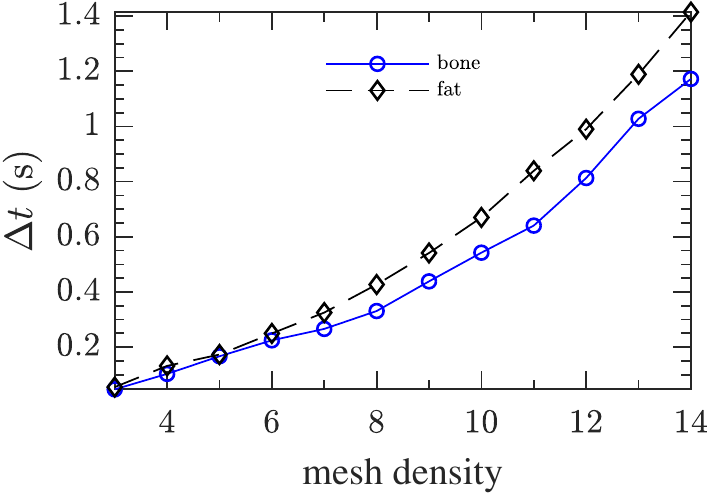}
        \end{center}
        \subcaption{The average GMRES time per iteration, i.e. $\Delta t$, at different mesh densities.}
        \label{fig:gmres_deltat}
    \end{subfigure}
    \caption{The convergence performance of the GMRES solver with tolerance $10^{-5}$ at different mesh densities. The mesh density is defined as the number of elements per wavelength.}
\end{figure}

In order to compare BEM and analytical results, the \ltwonorm of the difference in the total pressure, $p_{\mathrm{tot,err}}~=~p_{\mathrm{tot}} - p_{\mathrm{tot,an}}$ was calculated where $p_{\mathrm{tot}}$ is the computed pressure in the field by the BEM formulation and $p_{\mathrm{tot,an}}$ is the analytical total pressure. The computational grid is a uniform $200\times200$ grid spanning from $-3R$ to $3R$, where $R$ is the radius of the sphere, through the centre of the sphere. The predictions of the total field becomes more accurate, in the sense of $l_{\scriptscriptstyle 2}-\mathrm{norm}$, as the mesh density increases, see \figs{\ref{fig:norms_error}}. Nevertheless, with rather coarse surface meshes of $6$ elements per wavelength, the error is below $3\%$. This choice permits the prediction of the total field relatively accurately and fast-- the computation time per iteration (denoted by $\Delta t$) is about $0.2~\mathrm{s}$ for both fat and bone spheres, see \figs{\ref{fig:gmres_iters}}{\ref{fig:gmres_deltat}}. Bone is a high contrast material with $Z_{\mathrm{int}}/Z_{\mathrm{ext}}=5.2$ where $Z_{\mathrm{int}}$ and $Z_{\mathrm{ext}}$ are the characteristic specific acoustic impedances of the interior and exterior domains. For such materials, making the surface mesh finer improves the convergence rate of the iterative solver, as shown in \figs{\ref{fig:gmres_iters}}{\ref{fig:gmres_deltat}}. However, this comes at the expense of longer computation time per iteration because the matvec operations (the product of a matrix by a vector) of larger matrices need to be carried out. This improvement is negligible when the material contrast is small, \ie fat with $Z_{\mathrm{int}}/Z_{\mathrm{ext}}=0.86$.

\begin{figure}[t!]
    \begin{center}
        \includegraphics{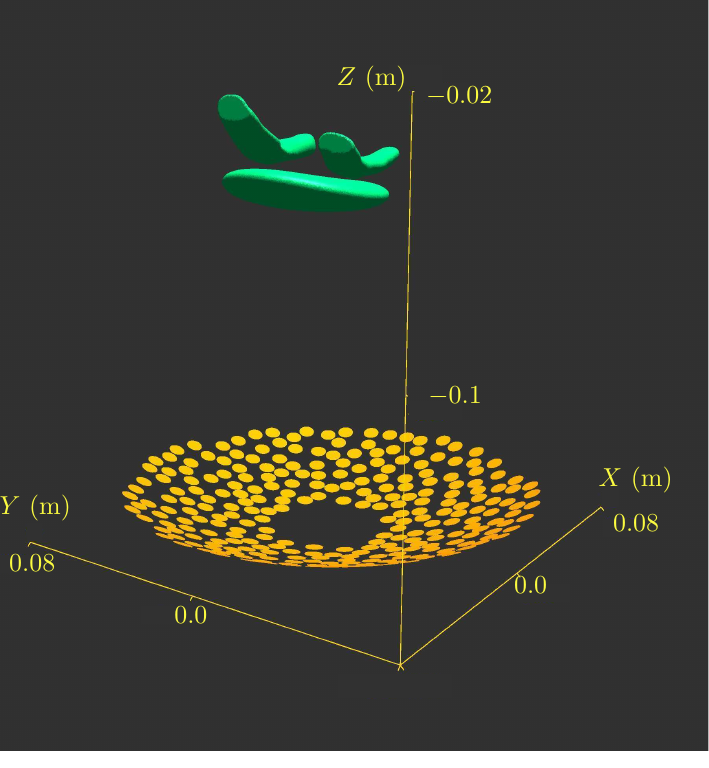}
    \end{center}
    \caption{Schematic diagram of the HIFU array, the abdominal fat layer and the ribcage. The intersection of visualisation axes is [-0.08,-0.08,-0.18].}
    \label{fig:hifu_ribs}
\end{figure}

\subsection{Anatomical model problems}\label{sec:hifu_ribs_kidney}

The results presented here are calculated for an incident field generated by a HIFU array transducer. The radius of the curvature of the array is 18 cm, the f-number is 1.18 and the focal point is located at the global origin by convention. The array features a pseudo-random spatial distribution of 256 piston elements, each of 6 mm diameter and facing towards the focal point. The details of calculating the HIFU incident field are presented in previous works, e.g. see \citep{elwin2015}, and will not be repeated here.

The first anatomical example involves a clinically relevant scenario of targeting a tumour of the liver located at an intercostal space 3.5 cm behind ribs 10 and 11, on the right side, see \figs{\ref{fig:hifu_ribs}}. The scattering objects (domains) are i) an abdominal fat layer located in the vicinity of the anterior of the ribcage, and ii) two human ribs. These two domains are immersed in an infinite exterior domain possessing physical properties representative of the human liver. Ribs generally consist of a cancellous bone core enclosed by a shell of cortical bone, the thickness of which varies with gender and age. Examining micro CT scans of ribs, it is apparent that the majority of the volume of a rib consists of cancellous bone \citep{perz2015,holcombe2018}. Thus, the physical properties of ribs used in the numerical simulations in this paper correspond to those of cancellous bone. The acoustic properties of all tissue media are listed in \Tab \ref{tab:physical_properties}. The fat layer was assumed to possess an ellipsoidal cross section with a thickness of 10 mm close to the axis of the transducer. The distance between the surface of the fat layer and the ribs vary from 1 mm to 15 mm.

\begin{figure}[t!]
    \begin{subfigure}[b]{0.45\columnwidth}
       \begin{center}
            \includegraphics{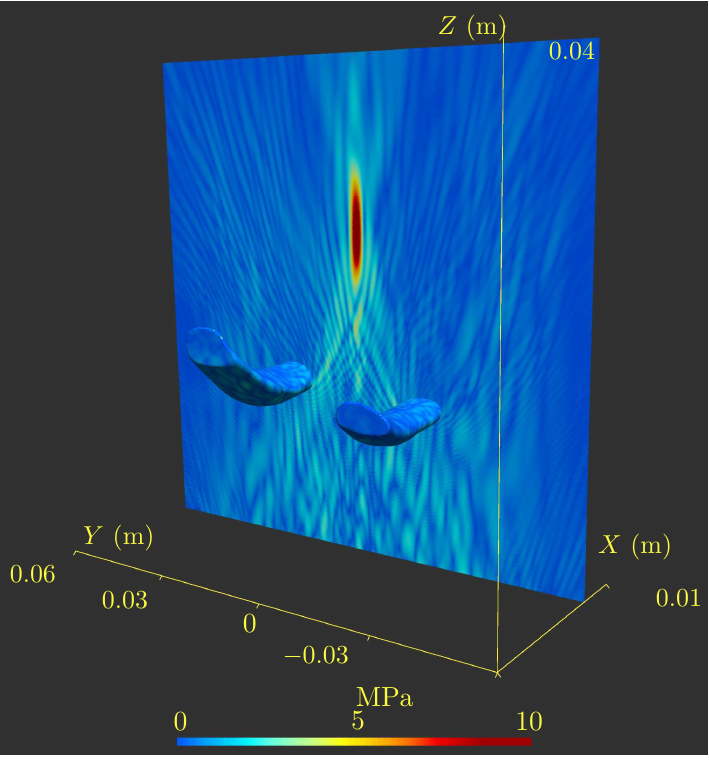}
        \end{center}
        \subcaption{Two ribs without a fat layer.}
        \label{fig:ptot_ribs}
    \end{subfigure}
    \hspace{5mm}
    \begin{subfigure}[b]{0.45\columnwidth}
        \begin{center}
            \includegraphics{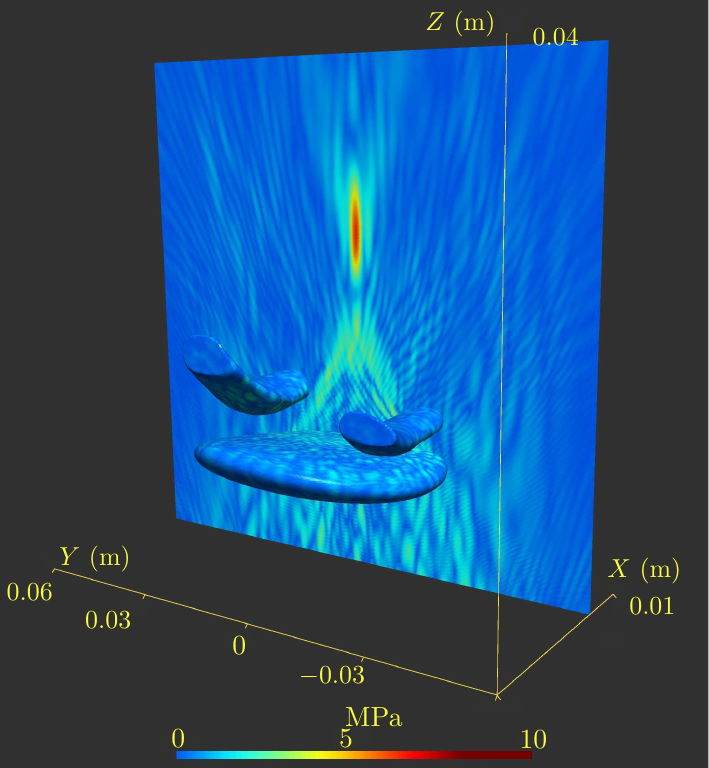}
        \end{center}
        \subcaption{Two ribs and a fat layer.}
        \label{fig:ptot_fat_ribs}
    \end{subfigure}
    \caption{Calculated absolute value of $p_\mathrm{tot}$ for the case of two ribs (a) without a fat layer and (b) with a fat layer.}
\end{figure}

The surface meshes of domain $j$ are made of triangles with an average element size of $\lambda/4$ where $\lambda~=~\min{(\lambda_0,\lambda_j)}, ~ j=1,2$. This resulted in i) 44924 nodes (89848 dofs) in the fat domain, and ii) 30088 nodes (60176 dofs) in the ribs domain.

\begin{figure}
    \begin{subfigure}[b]{0.45\columnwidth}
        \begin{center}
            \includegraphics{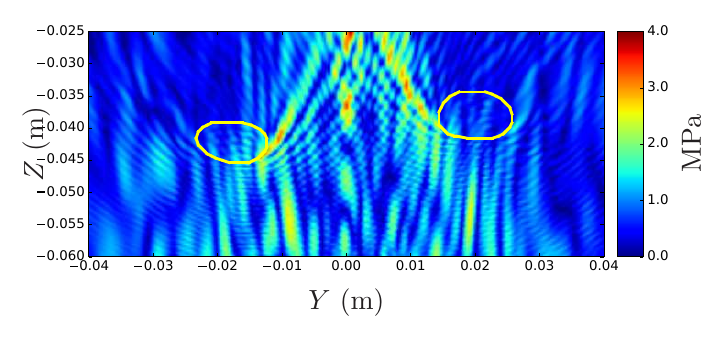}
        \end{center}
        \subcaption{Two ribs without a fat layer. }
        \label{fig:ptot_ribs_2d}
    \end{subfigure}
    \hspace{5mm}
    \begin{subfigure}[b]{0.45\columnwidth}   
        \begin{center}
            \includegraphics{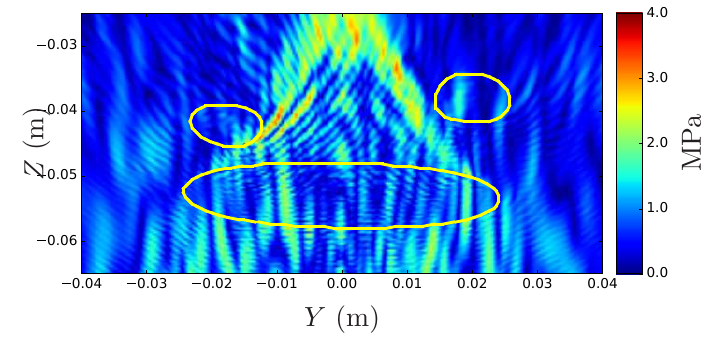}
        \end{center}
        \subcaption{Two ribs and a fat layer. }
        \label{fig:ptot_fat_ribs_2d}
    \end{subfigure}
    \caption{Calculated absolute value of $p_{\mathrm{tot}}$ at plane $Z=0$, for the case of two ribs (a) without a fat layer and (b) with a fat layer. The colour bar limit was set to $4 \mpa$ for a better visualisation of the interior fields. Contours specify the boundary of the ribs and the fat layer. }
\end{figure}

\begin{figure}
    \begin{subfigure}[b]{0.45\columnwidth}
        \begin{center}
            \includegraphics{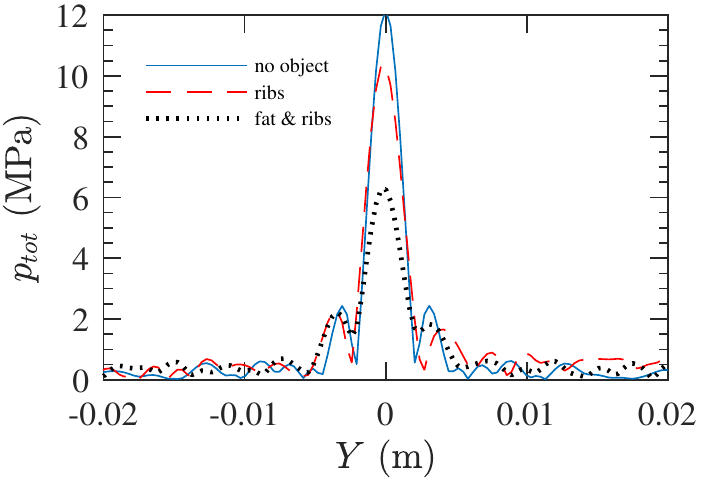}
        \end{center}
        \subcaption{}
    \label{fig:ptot_yaxis_fat_ribs}
    \end{subfigure}
    \hspace{5mm}
    \begin{subfigure}[b]{0.45\columnwidth}
        \begin{center}
            \includegraphics{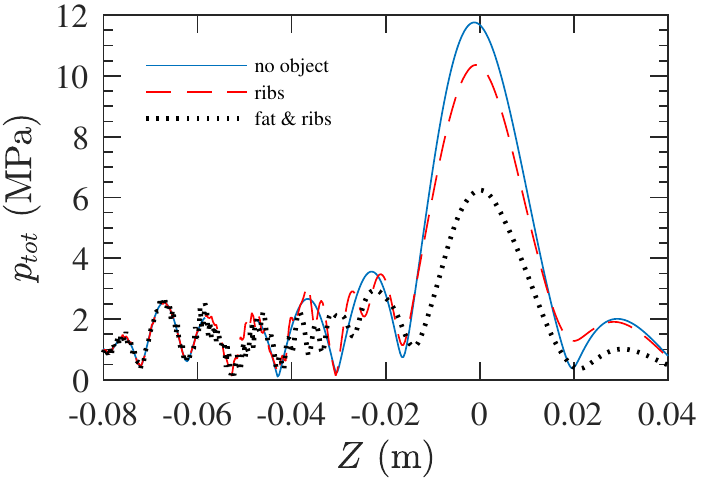}
        \end{center}
    \subcaption{}
    \label{fig:ptot_zaxis_fat_ribs}
    \end{subfigure}
    \caption{The absolute value of $p_{\mathrm{tot}}$ along (a) $Y$ axis at $X=0,Z=0$, (b) along $Z$ axis at $X=0,Y=0$.}
\end{figure}

In the first instance, a simulation was carried out omitting the fat layer. The GMRES solver (with the same tolerance and restarts as before) solved the preconditioned equations for the ribs model in 74 iterations and 2 minutes. Adding the fat layer to the model increased the number of iterations to 82 and the runtime to just under 9 minutes. Here, $R_\epsilon~=10^{-4}$ m was used to calculate the OSRC preconditioner, \eq{\ref{eq:OSRC_DtN_NtD_maps}}. The computation of the discrete matrices with the standard $\mathcal{H}$-matrix compression technique took 80 and 465 minutes for the ribs only and fat and ribs problems, respectively. The simulation results are depicted in \figs{\ref{fig:ptot_ribs}}{} to {\ref{fig:ptot_fat_ribs_2d}}. It can be observed that the presence of the fat layer leads to a drop in the peak pressure magnitude at the focus. Another interesting observation is the formation of a pre-focal high-pressure area caused by the constructive interference of the diffracted and transmitted waves through the fat layer. The significance of this observation and its clinical ramifications will however depend on the size, geometry and location of the fat layer with respect to the tumour and other tissue heterogeneities. The patient specificity of such features reinforces the need for treatment planning using validated numerical models, prior to a HIFU intervention, so that the potential impact of pre-focal heating may be gauged and subsequently mitigated. \figs{\ref{fig:ptot_yaxis_fat_ribs}}{\ref{fig:ptot_zaxis_fat_ribs}} show the axial and lateral waveforms through the focus (\ie along the $Z$ and $Y$ axes), respectively. The peak pressure at focus is reduced by almost $15\%$ when accounting for transmission and scattering by ribs. Augmenting the model by including the fat layer, the peak pressure at focus decreases by about $50\%$.

The second anatomical example involves another clinically relevant scenario using HIFU to ablate a tumour of the kidney. The apex of the spherical section array is positioned at [-0.13, 0.13, 0] m and the axis of transducer is colinear with the [1,-1,0] vector, thus resulting in the geometric focus of the array being inside the kidney. The scattering domains are i) a perinephric fat layer enclosing the kidney, and ii) a human kidney model. These two domains are immersed in an infinite exterior domain possessing physical properties of water. The physical properties of all domains are shown in \Tab \ref{tab:physical_properties}. The thickness of the perinephric fat layer is about 10 mm and is placed 1 mm away from the kidney (constant gap).

\begin{figure}[tbh]
        \begin{subfigure}[b]{0.4\columnwidth}
            \begin{center}
                \includegraphics{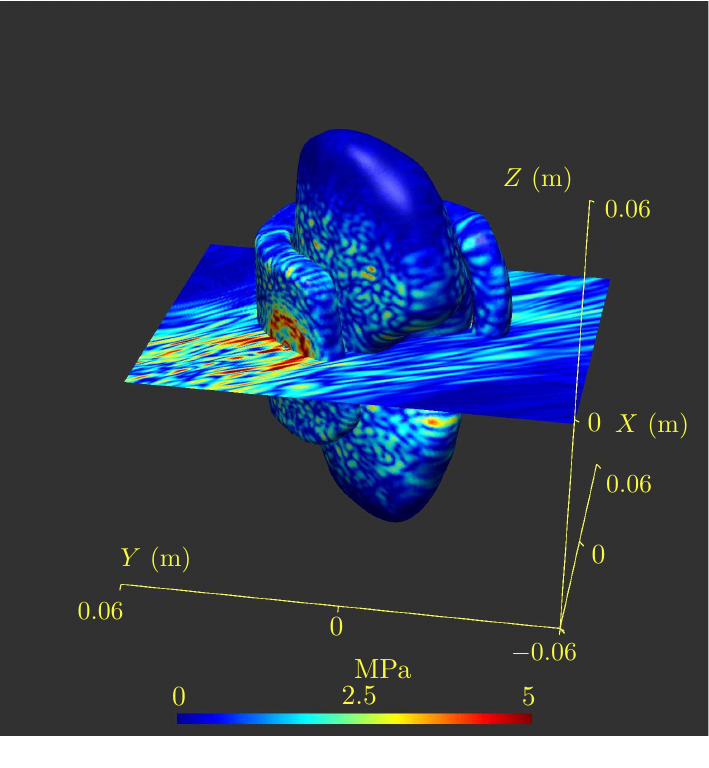}
            \end{center}
            \subcaption{}
            \label{fig:ptot_fat_kidney}
        \end{subfigure}
        \hspace{5mm}
        \begin{subfigure}[b]{0.6\columnwidth}
            \begin{center}
                \includegraphics{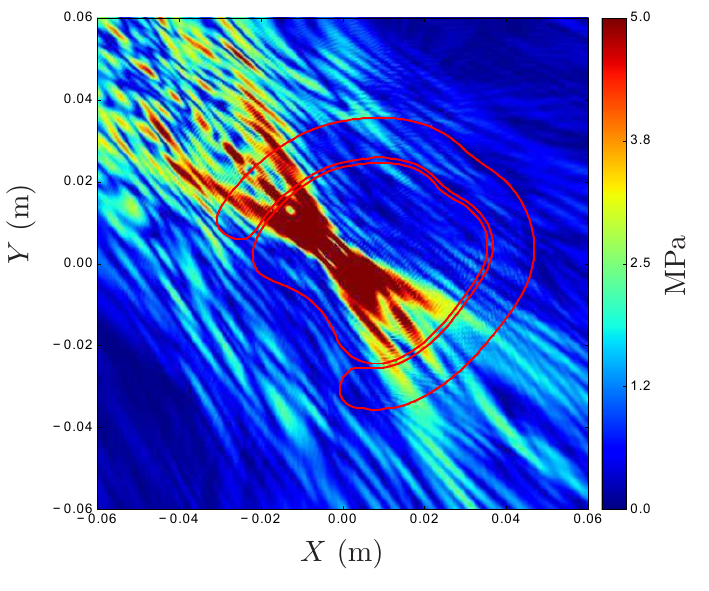}
            \end{center}
            \subcaption{}
            \label{fig:ptot_zplane_fat_kidney}    
        \end{subfigure}
    \caption{Calculated absolute value of $p_{\mathrm{tot}}$ for the case of a perinephric fat layer and the kidney: (a) in 3D, (b) at $Z=0$ plane.}
\end{figure}

\begin{figure}[t!]
    \begin{center}
        \includegraphics{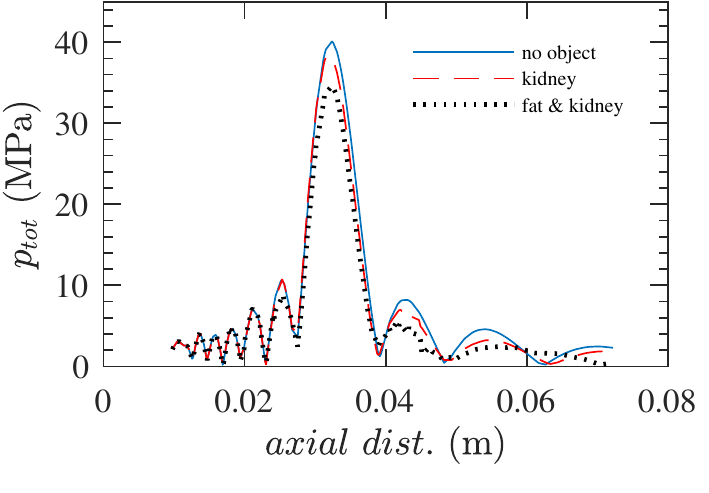}
    \end{center}
    \caption{The absolute value of $p_{\mathrm{tot}}$ at $Z=0$ plane along the axis of propagation, \ie [1,-1,0] vector. The axial distances are determined from the centre of the array located at [-0.13,-0.13,0].}
    \label{fig:ptot_wave_axis_fat_kidney}
    \end{figure}

Similar to the previous examples, the triangular surface meshes have an average element size of $\lambda/4$. The fat and kidney domains have 47766 and 65870 nodes (95532 and 131740 dofs), respectively. The $R_\epsilon$ parameter of the OSRC preconditioner was set to $10^{-3}$ m. The preconditioned equations were solved in 69 iterations which took about 35 minutes. Simulations are shown in \figs{\ref{fig:ptot_fat_kidney}}{\ref{fig:ptot_zplane_fat_kidney}}. Considering that the material contrasts in both domains are small, \ie $Z_{\mathrm{fat}}/Z_{\mathrm{water}}~=~0.86$ and $Z_{\mathrm{kidney}}/Z_{\mathrm{water}}~=~1.10$, the aberration of the HIFU beam is negligible. This is better displayed in \figs{\ref{fig:ptot_wave_axis_fat_kidney}} which shows the magnitude of the axial pressure for different scenarios. The peak pressure at the focus is reduced by $15\%$ due to the presence of the perinephric fat layer and the kidney. Eliminating the perinephric fat layer and keeping the kidney, the peak pressure at the focus only reduces by $5\%$. This problem was solved in 45 iterations and 12 minutes. The computation time to discretise and populate the matrix with $\mathcal{H}$-matrix compression is about 20 and 40 hours for the kidney only and kidney and perinephric fat models, respectively.

\section{Conclusion}
An innovative fast multiple-domain BEM formulation for solving the Helmholtz transmission problem was developed in this paper. The formulation uses a novel OSRC preconditioner and a variation of the multiple-domain PMCHWT. It was developed for HIFU treatment planning applications. The numerical simulations were performed to calculate the scattered field from i) a model of a human ribcage and an abdominal fat layer, and ii) a human kidney and perinephric fat layer, exposed to the acoustic field generated by a HIFU array transducer. The simulation results showed that the presence of tissue heterogeneities and strong scatterers such as bone can lead to substantial aberration of the focus, the formation of the pre-focal high-pressure area, and reduction in the peak pressure. The significance of these effects are patient specific which reinforces the need for treatment planning using validated numerical models prior to a HIFU intervention.

Furthermore, these results promote the application of the new OSRC block diagonal preconditioner and the multiple-domain BEM algorithm for fast and accurate patient-specific HIFU treatment planning for tumours of the abdomen. It was shown that the new preconditioner and formulation can still perform well when high contrast domains are present. Nevertheless, the computational performance of this solver can be further improved by speeding up the calculation of the \cald operators. This will be the subject of our forthcoming paper.  

\begin{table}[t]
    \caption{Physical properties of the domains \citep{matdatabase2018,duck1990}.}
    \scriptsize
    \begin{center}
    \begin{tabular*}{0.5\textwidth}{l | @{\extracolsep{\fill}} c|c|c|c}
    \hline
    domain & $\rho$                   & $c$                     & $\alpha_0$                        & $b$  \\
           & [$\mathrm{kg~m^{-3}}$]   & [$\mathrm{m~ s^{-1}}$]  & [$\mathrm{Np~ m^{-1}}$ at 1 $\mhz$] & [DL] \\
    \hline
    ribs   &           1178             &   2117.5                &    47                          & 1    \\
    fat    &           917              &   1412                  &    9.3                         & 1    \\
    kidney &           1066             &   1554.3                &    2.8067                      & 1    \\
    liver  &           1070             &   1640                  &    7                           & 1.30 \\
    water  &           1000             &   1500                  &    0.015                       & 2    \\
    \hline
    \end{tabular*}
    \end{center}
    \label{tab:physical_properties}
\end{table}

\ack{This work was supported by the EPSRC [Grant No. EP/P012434/1] and CONICYT [FONDECYT No. 11160462].}

\appendix

\section{\cald operator}\label{sec:CaldOp}

Solving the Helmholtz transmission problem with the BEM requires to reformulate the Helmholtz equation in the volume to boundary integral equations at the interfaces of two domains. This can be achieved by the aid of representation formulae which determine the acoustic fields from surface potentials at the boundaries. The representation formula for the domain $j,~(j \neq 0)$ reads

\begin{align}
    \label{eq:rep_formula_gen}
    p^j~=~\mathsf{SL}_{k_j}^j \psi^j(p^j)~-~\mathsf{DL}_{k_j}^j \phi^j(p^j),
\end{align}

\noindent where $\phi^j(p^j)$ and $\psi^j(p^j)$ are the jump potentials, and $\mathsf{SL}_{k_j}^j$ and $\mathsf{DL}_{k_j}^j$ are the single layer and double layer potential integral operators, respectively, and are given by 

\begin{align}
    \label{eq:SLP}
    \left[\mathsf{SL}^j \psi(p^j)\right](\mathbf{x})~=~\int_{\partial\Omega^j} G_{k_j} \left(\mathbf{x,y}\right)\psi(\mathbf{y}) d\Gamma(\mathbf{y})~ \nonumber \\
    \text{for}~ \mathbf{x}\notin \partial\Omega^j,   
\end{align}

\begin{align}
    \label{eq:DLP}
    \left[\mathsf{DL}^j\phi(p^j)\right](\mathbf{x})~=~\int_{\partial\Omega^j} \dfrac{\partial G_{k_j} \left(\mathbf{x,y}\right)}{\partial \mathbf{n}^j(\mathbf{y})}\phi(\mathbf{y}) d\Gamma(\mathbf{y})~ \nonumber\\
    \text{for} ~\mathbf{x}\notin \partial\Omega^j ,
\end{align}

\noindent where $G_{k_j} \left(\mathbf{x,y}\right)$ denotes the Green$^{\scriptscriptstyle {'}}$s function for the Helmholtz equation given by

\begin{align}
    \label{eq:green_func}
    G_{k_j} \left(\mathbf{x,y}\right)~=~\dfrac{e^{i k_j |\mathbf{x-y}|}}{4\pi |\mathbf{x-y}|}\nonumber \\
    \text{for} ~\mathbf{x,y}\in \Omega^j ~\text{and}~ \mathbf{x}\neq \mathbf{y}.
\end{align}

Taking the \emph{Cauchy} trace, $\gamma^{j,\pm}~:=~\begin{bmatrix}\gamma_D^{j,\pm} & \gamma_N^{j,\pm}\end{bmatrix}^T$, of the representation formula \eq{\ref{eq:rep_formula_gen}} produces

\begin{align}
    \label{eq:trace_repformula_rhs_1}
    &\gamma^{j,\pm}p^j~=~\gamma^{j,\pm}\mathsf{SL}_{k_j}^j(\psi^j(p^j))~-~ \nonumber \\
    &\gamma^{j,\pm} \mathsf{DL}_{k_j}^j(\phi^j(p^j))~= \nonumber \\
    &\begin{bmatrix}
    \gamma_D^{j,\pm} \mathsf{SL}_{k_j}^j(\psi^j(p^j)) ~-~\gamma_D^{j,\pm} \mathsf{DL}_{k_j}^j(\phi^j(p^j)) \\ \gamma_N^{j,\pm} \mathsf{SL}_{k_j}^j(\psi^j(p^j)) ~-~\gamma_N^{j,\pm} \mathsf{DL}_{k_j}^j(\phi^j(p^j))
    \end{bmatrix}.
\end{align}

In order to further simplify this, we need to use the jump relations and their regularity across the boundary. The jump relations read
    
\begin{align}
    \label{eq:jump_relations}
    & \gamma_D^{j,\pm} \mathsf{SL}_{k_j}^j(\psi^j(p^j)) ~=~ \mathsf{V}^j ~\psi^j(p^j), \nonumber \\
    & \gamma_D^{j,\pm} \mathsf{DL}_{k_j}^j(\phi^j(p^j)) ~=~ \mathsf{K}^j ~\phi^j(p^j)~\pm~\frac{1}{2}\phi^j(p^j), \nonumber \\
    & \gamma_N^{j,\pm} \mathsf{SL}_{k_j}^j(\psi^j(p^j)) ~=~ \mathsf{K^{'}}^j \psi^j(p^j)~\mp~\frac{1}{2}\psi^j(p^j), \nonumber \\
    & \gamma_N^{j,\pm} \mathsf{DL}_{k_j}^j(\phi^j(p^j)) ~=~ -\mathsf{W}^j \phi^j(p^j), \nonumber  \\
\end{align}

\noindent where $\mathsf{V}^j$, $\mathsf{K}^j$, $\mathsf{K^{'}}^j$, $\mathsf{W}^j$ are the single layer, double layer, adjoint double layer, and hypersingular boundary integral operators defined as follows

\begin{align}
    \label{eq:SLP}
    \left[\mathsf{V}^j \psi(p^j)\right](\mathbf{x})~=~\int_{\partial\Omega^j} G_{k_j} \left(\mathbf{x,y}\right)\psi(\mathbf{y}) d\Gamma(\mathbf{y})~\nonumber \\
    \text{for}~ \mathbf{x}\in \partial\Omega^j,   
\end{align}

\begin{align}
    \label{eq:DLP}
    \left[\mathsf{K}^j\psi(p^j)\right](\mathbf{x})~=~\int_{\partial\Omega^j} \dfrac{\partial G_{k_j} \left(\mathbf{x,y}\right)}{\partial \mathbf{n}^j(\mathbf{y})}\psi(\mathbf{y}) d\Gamma(\mathbf{y})~ \nonumber\\
    \text{for}~ \mathbf{x}\in \partial\Omega^j,  
\end{align}

\begin{align}
    \label{eq:ADJDLP}
   &\left[\mathsf{K^{'}}^j\psi(p^j)\right](\mathbf{x})~=~\nonumber\\
   &\dfrac{\partial}{\partial \mathbf{n}^j(\mathbf{x})} \int_{\partial\Omega^j}  G_{k_j} \left(\mathbf{x,y}\right)\psi(\mathbf{y}) d\Gamma(\mathbf{y})~, \text{for}~ \mathbf{x}\in \partial\Omega^j,
\end{align}

\begin{align}
    \label{eq:HYP}
    & \left[\mathsf{W}^j\psi(p^j)\right](\mathbf{x}) ~=~ \nonumber\\ 
    & - \dfrac{\partial}{\partial \mathbf{n}^j(\mathbf{x})} \int_{\partial\Omega^j} \dfrac{\partial G_{k_j} \left(\mathbf{x,y}\right)}{\partial \mathbf{n}^j(\mathbf{y})} & \psi(\mathbf{y}) d\Gamma(\mathbf{y})~ \nonumber\\
    &&\text{for}~ \mathbf{x}\in \partial\Omega^j .
\end{align}

Inserting the jump relations into \eq{\ref{eq:trace_repformula_rhs_1}} results in

\begin{align}
& \begin{bmatrix}
\mathsf{V}^j ~\psi^j(p^j)~-~\mathsf{K}^j ~\phi^j(p^j)~\mp~\frac{1}{2}\phi^j(p^j) \\
\mathsf{K^{'}}^j \psi^j(p^j)~\mp~\frac{1}{2}\psi^j(p^j)~+~\mathsf{W}^j \phi^j(p^j)
\end{bmatrix}~=~ \nonumber \\
& \begin{bmatrix}
-~\mathsf{K}^j~\mp~\frac{1}{2}\mathsf{Id} & \mathsf{V}^j \\
\mathsf{W}^j & \mathsf{K^{'}}^j~\mp~\frac{1}{2}\mathsf{Id}
\end{bmatrix} \begin{bmatrix}\phi^j(p^j) \\\psi^j(p^j)\end{bmatrix},
\end{align}

\noindent which taking the identity matrix out gives the \emph{\cald operator} defined as follows

\begin{align}
\mathsf{A}_{k_j}^j~:=~\begin{bmatrix}
-~\mathsf{K}^j  & \mathsf{V}^j  \\
\mathsf{W}^j &  \mathsf{K^{'}}^j
\end{bmatrix},
\end{align}

\noindent thus, \eq{\ref{eq:trace_repformula_rhs_1}} becomes

\begin{align}
\label{eq:trace_repformula_rhs_2}
\gamma^{j,\pm}p^j~=~\left( \mp\frac{1}{2}\mathsf{Id}~+~\mathsf{A}_{k_j}^j \right)\begin{bmatrix}\phi^j(p^j) \\\psi^j(p^j)\end{bmatrix}.
\end{align}

\noindent For the PMCHWT formulation, the interior jump potentials read 

\begin{subequations}
    \label{eq:Form2_jump_potentials}
    \begin{align}
    & \phi^j(p^{j})~=~ \gamma_D^{j,-}p^j, \\
    & \psi^j(p^{j})~=~  \gamma_N^{j,-}p^j. 
    \end{align}
\end{subequations}

\noindent Taking \eq{\ref{eq:trace_repformula_rhs_2}}{\ref{eq:Form2_jump_potentials}} into consideration, the \cald representation of the interior Helmholtz problem in domain $\Omega^j$ reads

\begin{align}
    \label{eq:Cald_formula_int}
    \gamma^{j,-}p^j~=~\left( \frac{1}{2}\mathsf{Id}~+~\mathsf{A}_{k_j}^j \right)\gamma^{j,-}p^j.
\end{align}

\noindent With regards the exterior domain, we define the total scattered field in $\Omega^0$ with $p_s$ and the total exterior pressure as $p_{\mathrm{tot}}~=~p_s + p_{\mathrm{inc}}$. Incorporating these definitions, the exterior jump potentials for PMCHWT formulation become $\phi^0(p^0)~=~ -\gamma_D^{+} p_{\mathrm{tot}}$ and $\psi^0(p^0)~=~ -\gamma_N^{+} p_{\mathrm{tot}}$, and $\gamma^{0,+}p~=~\gamma^{+} p_s$ where the \emph{Cauchy} traces $\gamma^+$ are evaluated at $\bigcup_{j=1}^n \partial\Omega^j$. Substituting these equations into \eq{\ref{eq:trace_repformula_rhs_2}} yields the \cald formulation of the exterior Helmholtz problem as follows, one should note that the \cald boundary operator is evaluated with the exterior wavenumber $k_0$,

\begin{align}
    \label{eq:Cald_formula_ext0}
    \gamma^{+}p_s~=~\left( \frac{1}{2}\mathsf{Id}~-~\mathsf{A}_{k_0} \right)\gamma^{+}p_{\mathrm{tot}}.
\end{align}

\bibliography{references}

\end{document}